\documentclass[reprint,superscriptaddress,showpacs,secnumarabic,amssymb, nobibnotes, aps, prl,floatfix]{revtex4-1}
\pdfoutput=1

\usepackage{graphicx}
\usepackage{epstopdf}
\usepackage{color}
\usepackage{amsmath}
\begin{document}

\title{Longitudinal target-spin asymmetries for deeply virtual Compton scattering}
\newcommand*{\UCONN}{University of Connecticut, Storrs, Connecticut 06269}
\newcommand*{\UCONNindex}{1}
\affiliation{\UCONN}
\newcommand*{\SACLAY}{CEA, Centre de Saclay, Irfu/Service de Physique Nucl\'eaire, 91191 Gif-sur-Yvette, France}
\newcommand*{\SACLAYindex}{2}
\affiliation{\SACLAY}
\newcommand*{\FAIRFIELD}{Fairfield University, Fairfield, Connecticut 06824}
\newcommand*{\FAIRFIELDindex}{3}
\affiliation{\FAIRFIELD}
\newcommand*{\INFNFR}{INFN, Laboratori Nazionali di Frascati, 00044 Frascati, Italy}
\newcommand*{\INFNFRindex}{4}
\affiliation{\INFNFR}
\newcommand*{\ORSAY}{Institut de Physique Nucl\'eaire Orsay, 91406 Orsay, France}
\newcommand*{\ORSAYindex}{5}
\affiliation{\ORSAY}
\newcommand*{\GLASGOW}{University of Glasgow, Glasgow G12 8QQ, United Kingdom}
\newcommand*{\GLASGOWindex}{6}
\affiliation{\GLASGOW}
\newcommand*{\EDINBURGH}{Edinburgh University, Edinburgh EH9 3JZ, United Kingdom}
\newcommand*{\EDINBURGHindex}{7}
\affiliation{\EDINBURGH}
\newcommand*{\ODU}{Old Dominion University, Norfolk, Virginia 23529}
\newcommand*{\ODUindex}{8}
\affiliation{\ODU}
\newcommand*{\JLAB}{Thomas Jefferson National Accelerator Facility, Newport News, Virginia 23606}
\newcommand*{\JLABindex}{9}
\affiliation{\JLAB}
\newcommand*{\INFNGE}{INFN, Sezione di Genova, 16146 Genova, Italy}
\newcommand*{\INFNGEindex}{10}
\affiliation{\INFNGE}
\newcommand*{\ITEP}{Institute of Theoretical and Experimental Physics, Moscow, 117259, Russia}
\newcommand*{\ITEPindex}{11}
\affiliation{\ITEP}
\newcommand*{\FIU}{Florida International University, Miami, Florida 33199}
\newcommand*{\FIUindex}{12}
\affiliation{\FIU}
\newcommand*{\WM}{College of William and Mary, Williamsburg, Virginia 23187-8795}
\newcommand*{\WMindex}{13}
\affiliation{\WM}
\newcommand*{\GWUI}{The George Washington University, Washington, D.C. 20052}
\newcommand*{\GWUIindex}{14}
\affiliation{\GWUI}
\newcommand*{\UTFSM}{Universidad T\'{e}cnica Federico Santa Mar\'{i}a, Casilla 110-V Valpara\'{i}so, Chile}
\newcommand*{\UTFSMindex}{15}
\affiliation{\UTFSM}
\newcommand*{\OHIOU}{Ohio University, Athens, Ohio  45701}
\newcommand*{\OHIOUindex}{16}
\affiliation{\OHIOU}
\newcommand*{\INFNRO}{INFN, Sezione di Roma Tor Vergata, 00133 Roma, Italy}
\newcommand*{\INFNROindex}{17}
\affiliation{\INFNRO}
\newcommand*{\ISU}{Idaho State University, Pocatello, Idaho 83209}
\newcommand*{\ISUindex}{18}
\affiliation{\ISU}
\newcommand*{\INFNFE}{INFN, Sezione di Ferrara, 44100 Ferrara, Italy}
\newcommand*{\INFNFEindex}{19}
\affiliation{\INFNFE}
\newcommand*{\VIRGINIA}{University of Virginia, Charlottesville, Virginia 22901}
\newcommand*{\VIRGINIAindex}{20}
\affiliation{\VIRGINIA}
\newcommand*{\FSU}{Florida State University, Tallahassee, Florida 32306}
\newcommand*{\FSUindex}{21}
\affiliation{\FSU}
\newcommand*{\ROMAII}{Universit\`a di Roma Tor Vergata, 00133 Roma, Italy}
\newcommand*{\ROMAIIindex}{22}
\affiliation{\ROMAII}
\newcommand*{\YEREVAN}{Yerevan Physics Institute, 375036 Yerevan, Armenia}
\newcommand*{\YEREVANindex}{23}
\affiliation{\YEREVAN}
\newcommand*{\SCAROLINA}{University of South Carolina, Columbia, South Carolina 29208}
\newcommand*{\SCAROLINAindex}{24}
\affiliation{\SCAROLINA}
\newcommand*{\CNU}{Christopher Newport University, Newport News, Virginia 23606}
\newcommand*{\CNUindex}{25}
\affiliation{\CNU}
\newcommand*{\ANL}{Argonne National Laboratory, Argonne, Illinois 60439}
\newcommand*{\ANLindex}{26}
\affiliation{\ANL}
\newcommand*{\MSU}{Skobeltsyn Institute of Nuclear Physics, Lomonosov Moscow State University, 119234 Moscow, Russia}
\newcommand*{\MSUindex}{27}
\affiliation{\MSU}
\newcommand*{\INFNTUR}{INFN, Sezione di Torino, Torino, Italy}
\newcommand*{\INFNTURindex}{28}
\affiliation{\INFNTUR}
\newcommand*{\JMU}{James Madison University, Harrisonburg, Virginia 22807}
\newcommand*{\JMUindex}{29}
\affiliation{\JMU}
\newcommand*{\RPI}{Rensselaer Polytechnic Institute, Troy, New York 12180-3590}
\newcommand*{\RPIindex}{30}
\affiliation{\RPI}
\newcommand*{\UNH}{University of New Hampshire, Durham, New Hampshire 03824-3568}
\newcommand*{\UNHindex}{31}
\affiliation{\UNH}
\newcommand*{\TEMPLE}{Temple University,  Philadelphia, Pennsylvania 19122 }
\newcommand*{\TEMPLEindex}{32}
\affiliation{\TEMPLE}
\newcommand*{\NSU}{Norfolk State University, Norfolk, Virginia 23504}
\newcommand*{\NSUindex}{33}
\affiliation{\NSU}
\newcommand*{\KNU}{Kyungpook National University, Daegu 702-701, Republic of Korea}
\newcommand*{\KNUindex}{34}
\affiliation{\KNU}
\newcommand*{\CUA}{Catholic University of America, Washington, D.C. 20064}
\newcommand*{\CUAindex}{35}
\affiliation{\CUA}
\newcommand*{\ASU}{Arizona State University, Tempe, Arizona 85287-1504}
\newcommand*{\ASUindex}{36}
\affiliation{\ASU}
\newcommand*{\CSUDH}{California State University, Dominguez Hills, Carson, California 90747}
\newcommand*{\CSUDHindex}{37}
\affiliation{\CSUDH}
\newcommand*{\CMU}{Carnegie Mellon University, Pittsburgh, Pennsylvania 15213}
\newcommand*{\CMUindex}{38}
\affiliation{\CMU}
\newcommand*{\Genova}{Universit\`a di Genova, 16146 Genova, Italy}
\newcommand*{\Genovaindex}{39}
\affiliation{\Genova}
\newcommand*{\LPSC}{LPSC, Universit\'{e} Grenoble-Alps, CNRS/IN2P3, Grenoble, France}
\newcommand*{\LPSCindex}{40}
\affiliation{\LPSC}
\newcommand*{\CANISIUS}{Canisius College, Buffalo, New York 14208}
\newcommand*{\CANISIUSindex}{41}
\affiliation{\CANISIUS}
\newcommand*{\NOWUK}{University of Kentucky, Lexington, Kentucky 40506}
\newcommand*{\NOWJLAB}{Thomas Jefferson National Accelerator Facility, Newport News, Virginia 23606}
\newcommand*{\NOWUCONN}{University of Connecticut, Storrs, Connecticut 06269}
\newcommand*{\NOWODU}{Old Dominion University, Norfolk, Virginia 23529}
\newcommand*{\NOWGLASGOW}{University of Glasgow, Glasgow G12 8QQ, United Kingdom}
\newcommand*{\NOWINFNGE}{INFN, Sezione di Genova, 16146 Genova, Italy}
\newcommand*{\NOWUMASS}{University of Massachusetts, Amherst, Massachusetts  01003}
 %%%%%%%%%%%%%%% END OF Latex Macros for institute addresses  %%%%%%%%%%%%%%%%%%%%%%%%% 
\author{E.~Seder}
\affiliation{\UCONN}
\affiliation{\SACLAY}
\author{A.~Biselli}
\affiliation{\FAIRFIELD}
\author{S.~Pisano}
\affiliation{\INFNFR}
\affiliation{\ORSAY}
\author{S.~Niccolai}
\email[corresponding author: ]{silvia@jlab.org}
\affiliation{\ORSAY}
\author{G.D.~Smith}
\affiliation{\GLASGOW}
\affiliation{\EDINBURGH}
\author{K.~Joo}
\affiliation{\UCONN}
\author {K.~Adhikari}
\affiliation{\ODU}
\author {M.J.~Amaryan} 
\affiliation{\ODU}
\author {M.D.~Anderson} 
\affiliation{\GLASGOW}
\author {S.~Anefalos Pereira}
\affiliation{\INFNFR}
% ...

\collaboration{The CLAS Collaboration}
\noaffiliation

\date{\today}

\begin{abstract}
\noindent A measurement of the electroproduction of photons off protons in the deeply inelastic regime was performed at Jefferson Lab using a nearly 6-GeV electron beam, a longitudinally polarized proton target and the CEBAF Large Acceptance Spectrometer. 
Target-spin asymmetries for $ep\to e'p'\gamma$ events, which arise from the interference of the deeply virtual Compton scattering and the Bethe-Heitler processes, were extracted over the widest kinematics in $Q^2$, $x_B$, $t$ and $\phi$, for 166 four-dimensional bins. In the framework of Generalized Parton Distributions (GPDs), at leading twist the $t$ dependence of these asymmetries provides insight on the spatial distribution of the axial charge of the proton, which appears to be concentrated in its center. These results also bring important and necessary constraints for the existing parametrizations of chiral-even GPDs. 
\end{abstract}

\pacs{12.38.-t, 13.40.-f, 13.60.-r, 25.30.-c, 25.30.Rw, 25.30.Dh, 25.30.Fj}
%PACS: Chromodynamics, quantum; Electromagnetic interactions, Electron-induced nuclear reactions;
\maketitle

Nearly 60 years after Hofstadter's direct measurement of the finite size of the proton \cite{hofstadter}, the way the bulk properties of the nucleon, such as its mass and spin, are connected to the dynamics of its constituents is still a subject of intense research. Quantum Chromo-Dynamics (QCD), the fundamental theory of the strong interaction, is still unsolved for quarks confined in the nucleon. Therefore, phenomenological functions need to be used to connect experimental observables with the inner dynamics of the constituents of the nucleons, the partons. 
The Generalized Parton Distributions (GPDs), introduced two decades ago, have emerged as a universal tool to describe hadrons, and nucleons in particular, in terms of their elementary constituents, quarks and gluons \cite{muller,radyuskin,ji,diehl1,belitski,belitski-rad}. The GPDs combine and generalize the features of the form factors measured in elastic scattering and of the parton distribution functions obtained via deep inelastic scattering (DIS). In a reference frame in which the nucleon moves at the speed of light, the GPDs correlate the longitudinal momentum and the transverse position of partons in a given helicity state. They can also give access to the contribution to the nucleon spin from the orbital angular momentum of the quarks, via Ji's sum rule \cite{ji}.
At leading order in the QCD coupling constant $\alpha_s$ and at leading twist (i.e. neglecting quark-gluon interactions or higher-order quark loops), considering only quark GPDs and quark-helicity conserving quantities, there are four different GPDs for the nucleon: $H$, $E$, $\widetilde{H}$, $\widetilde{E}$, which can be measured in exclusive electroproduction reactions at high electron-momentum transfer.

\begin{figure}
\includegraphics[scale=0.5]{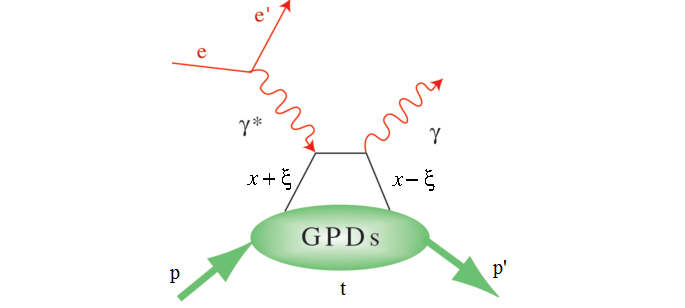}
\vspace{-0.4cm}
\caption{(Color online) The ``handbag'' diagram for the DVCS process on the proton $ep\to e'p'\gamma$. $t=(p-p')^2$ is the squared four-momentum transfer between the initial and final protons. $\xi$ is proportional to the Bjorken variable $x_B$ ($\xi\simeq\frac{x_B}{2-x_B}$, where $x_B=\frac{Q^2}{2M\nu}$, $M$ is the proton mass and $\nu=E_e-E_{e'}$). $x$ is not accessible experimentally in the DVCS process.}
\vspace{-0.4cm}
\label{fig1}
\end{figure}

Deeply virtual Compton scattering (DVCS) ($ep \to e'p'\gamma$, Fig.~\ref{fig1}) is the simplest process to access the GPDs of the proton. At high $\gamma^*$ virtuality $Q^2=-(e-e')^2$, and at leading twist, which is valid at small squared momentum transfer to the proton $-t$ relative to $Q^2$, this process corresponds to the absorption of a virtual photon by a quark carrying a fraction ($x+\xi$) of the longitudinal momentum of the proton with respect to its direction. The struck quark emits a real photon, as a result of which its final longitudinal momentum fraction is ($x-\xi$). The amplitude for DVCS can be factorized \cite{ji} into a hard-scattering part (calculable in perturbative QCD) and a non-perturbative part, representing the soft structure of the nucleon, parametrized by the GPDs that depend on the three kinematic variables $x$, $\xi$, and $t$. The definitions of the kinematic variables are in the caption of Fig.~\ref{fig1}. The Fourier transform, at $\xi=0$, of the $t$ dependence of a GPD provides the spatial distribution in the transverse plane for partons having a longitudinal momentum fraction $x$.

DVCS shares the same final state with the Bethe-Heitler (BH) process, where a real photon is emitted by either the incoming or the scattered electron. At the cross-section level BH is typically larger than DVCS, but information on the latter can be obtained by extracting the DVCS/BH interference term, and exploiting the fact that the amplitude from BH can be accurately computed. Using the invariance of the strong and electromagnetic interactions under parity and time reversal, it can be shown that a spin-dependent asymmetry, with respect to the spin of either the incoming electron or the target nucleon, of the $ep\to e'p'\gamma$ reaction at leading twist depends mainly on the DVCS/BH interference. Such spin asymmetries can then be connected to combinations of real and imaginary parts of Compton Form Factors (CFFs), defined as \cite{belitski}
\begin{equation}\label{cff_definition_0}
\Re{\rm e}{\cal F} = {\cal P}\int_{-1}^{1}dx\left[\frac{1}{x-\xi}\mp\frac{1}{x+\xi}\right]F(x,\xi,t) \nonumber 
\end{equation}
\begin{equation}\label{cff_definition}
\Im{\rm m}\mathcal{F} = \pi\left[F(\xi,\xi,t)\mp F(-\xi,\xi,t)\right],
\end{equation}

where $F$ represents any of the four GPDs, ${\cal P}$ is the principal value integral, and the top and bottom signs correspond, respectively, to the quark-helicity independent ($H$, $E$) and the quark-helicity dependent ($\widetilde{H}$, $\widetilde{E}$) GPDs.

Depending on the polarization observable extracted, different sensitivities to the four GPDs can be realized. For instance, the target-spin asymmetry for a longitudinally polarized proton target, denoted by $A_{UL}$, is sensitive to a combination of $\Im{\rm m}\widetilde{\cal H}$ and $\Im{\rm m}{\cal H}$. Conversely, the beam-spin asymmetry measured using a polarized beam is dominated by ${\Im{\rm m}\cal H}$.  While $H$ is connected to the distribution of the electric charge in the nucleon, the $\widetilde{H}$ GPD is related to the nucleon axial charge \cite{belitski-rad}, which expresses the probability that an axial particle (such as $W$, $Z$, $a_1$,...) couples to the nucleon, providing a bridge between the strong and the weak interactions. At leading twist $A_{UL}$ can be expressed as a function of the angle $\phi$ between the leptonic $(e \times e')$ and hadronic $(\gamma^* \times p')$ planes for each bin in $(Q^2,\xi,t)$ as \cite{belitski-rad}:
\begin{equation}\label{eq_aUL}
A_{UL}(\phi)\sim \frac{\alpha\sin\phi}{1+\beta\cos\phi},
\end{equation}
where the $\beta$ term arises mainly from the BH amplitude, while the GPDs appear in the DVCS/BH interference term $\alpha$ as a linear combination of the four imaginary parts of the CFFs. The coefficients of this sum, which are $(Q^2, \xi, t)$-dependent kinematic factors and the precisely-known electromagnetic form factors, enhance the contribution of $\Im{\rm m}\widetilde{\cal H}$ and, in a lesser way, of $\Im{\rm m}{\cal H}$ with respect to the other CFFs. 
Beyond the leading twist, i.e. when $-t\sim Q^2$, additional $\sin(n\phi)$ terms, with $n\geq 2$, appear in the numerator of Eq.~\ref{eq_aUL}.

After the first observations of a $\sin\phi$ dependence for $ep\to e'p'\gamma$ events --- a signature of the DVCS/BH interference --- in low-statistics beam-spin asymmetry measurements \cite{stepan,hermes_bsa}, various high-statistics DVCS experiments were performed. As of today, polarized and unpolarized cross sections measured at Jefferson Lab Hall A \cite{carlos} indicate, via a $Q^2$-scaling test, that the factorization and leading-twist approximations are valid already at relatively low $Q^2$ ($\sim 1-2$ (GeV/c)$^2$). High-statistics and wide-coverage beam-spin asymmetries measured in Hall B with CLAS \cite{fx} brought important constraints for the parametrization of the GPD $H$. 
Exploratory measurements of the longitudinal target-spin asymmetry were made by CLAS \cite{shifeng} and HERMES \cite{hermes}, but the low statistical precision of the data did not allow to map simultaneously its $Q^2$, $x_B$, $t$, and $\phi$ dependence. Therefore, unlike $H$, the GPD $\widetilde{H}$ has not yet been well constrained. 

This paper presents results of longitudinal target-spin asymmetries for DVCS/BH obtained, for the first time, over a large phase space and in four-dimensional bins in $Q^2$, $x_B$, $t$, and $\phi$. The data were taken in Hall B at Jefferson Lab in 2009, using a polarized electron beam with an average energy of 5.932 GeV that impinged on a solid, dynamically-polarized 1.5-cm-long ammonia target \cite{chris}. Protons in paramagnetically doped $^{14}$NH$_{3}$ were continuously polarized along the beam direction at 5T and 1K by microwave irradiation. A superconducting, split-coil magnet provided the uniform polarizing field for the target and at the same time focused the low-energy M{\o}ller electrons towards the beam line, away from the detectors. Periodically, data were taken on a $^{12}$C target, to allow unpolarized-background studies. 

The scattered electron, the recoil proton, and the photon were detected in the CEBAF Large Acceptance Spectrometer (CLAS) \cite{clas}, which allowed multiple-particle identification with a wide acceptance. A totally absorbing Faraday cup downstream of CLAS was used to determine the integrated beam charge passing through the target. The trigger, defined by the scattered electron, was provided by matching signals in the same sector for the Cherenkov counters and the electromagnetic calorimeters (EC). In offline analysis, energy cuts on the EC allowed for rejection of the negative-pion background. Protons, deflected by the magnetic field of the superconducting toroid, passed through three regions of drift chambers, for momentum measurement, and reached an array of scintillator paddles, for time-of-flight measurement and particle identification. Photons were detected by the EC for polar angles from 17${}^{\circ}$ to 43${}^{\circ}$ and by the Inner Calorimeter (IC) \cite{fx} from 4${}^{\circ}$ to 15${}^{\circ}$. 

\begin{figure} 
\includegraphics[width = 8.6cm]{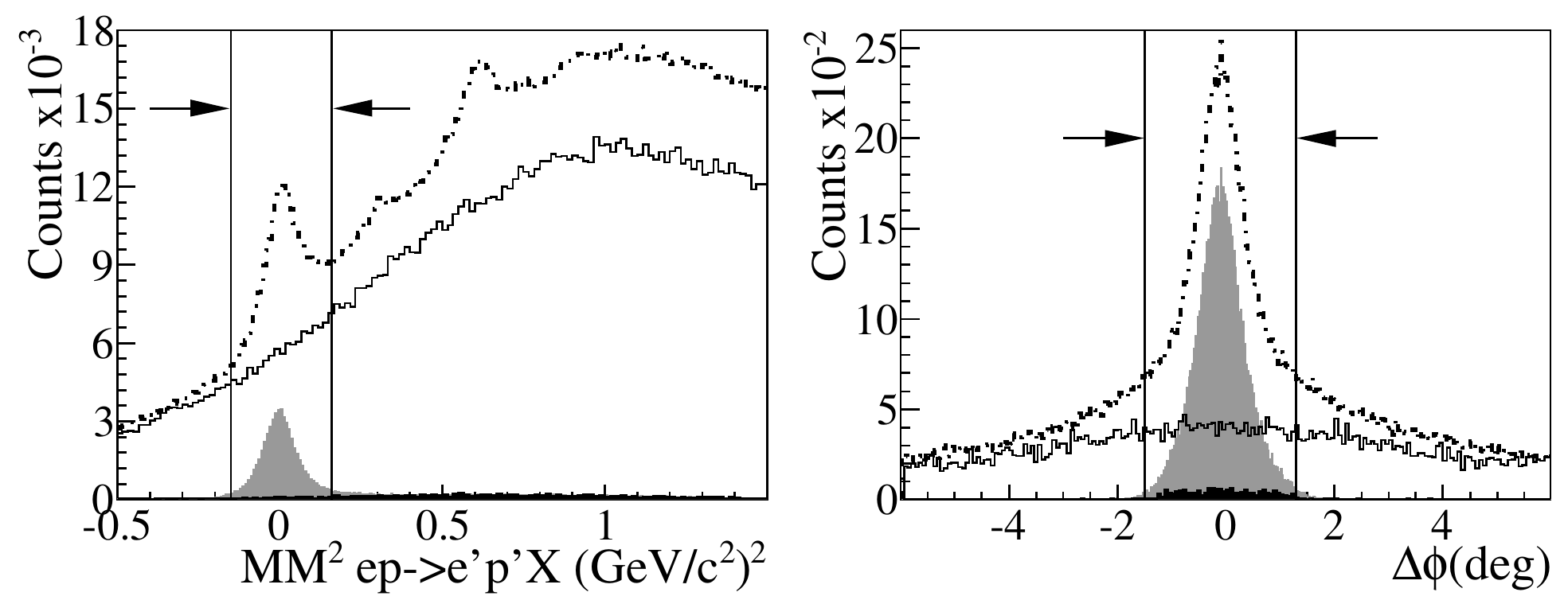}
\vspace{-0.7cm}
\caption{\label{fig2} Left: squared missing mass (MM) of $X$ in the $ep\to e'p'X$ reaction; right: $\Delta\phi$. The dot-dashed and solid lines show the events {\it before} exclusivity cuts for, respectively, $^{14}$NH$_{3}$ and $^{12}$C data; the gray and black shaded plots are the events {\it after} all exclusivity cuts but the one on the plotted variable for, respectively, $^{14}$NH$_{3}$ and $^{12}$C data. The lines and arrows show the limits of the cuts. The plots for $^{14}$NH$_{3}$ and $^{12}$C data are normalized to each other via their relative luminosities.}
\vspace{-0.4cm}
\end{figure}

Once all three DVCS/BH final-state particles ($ep\gamma$) were identified and their momenta and angles measured, channel-selection cuts were applied on the following four quantities: the missing mass of $X$ in the $ep \to e'p'X$ reaction, the missing transverse momentum $p_t(X)$ in the $ep \to e'p'\gamma X$ reaction, the cone angle $\theta_{\gamma X}$ between the measured and the kinematically reconstructed photon from $ep \to e'p'X$, and the difference $\Delta\phi$ between two ways to compute the angle $\phi$ (defining the hadronic plane using the directions of the proton and of either the real or the virtual photon). 
Figure \ref{fig2} shows, as examples, the effect of the cuts on the missing mass of $X$ in $ep \to e'p'X$ (left) and on $\Delta\phi$ (right). The gray and black shaded areas represent the events after all exclusivity cuts but the one on the plotted variable were applied for the $^{14}$NH$_{3}$ and $^{12}$C data, respectively. The black shaded areas, in particular, show how the cuts drastically reduce the effect of the nuclear background from $^{14}$N. The remaining  $^{14}$N contamination was evaluated using a dilution factor determined from the carbon data. This factor ($D_{f}\sim0.92$), which accounts for the fraction of $e'p'\gamma$ events originating from the polarized hydrogen relative to the total number of $e'p'\gamma$ events from all materials in the target, was applied to the final asymmetry (see Eq.~\ref{eq_asym} below). 
%%%%%%%%%%%%%%%%%%%%%%%%%%%%%%%%%%%%%%%%%%%%%%%%%%%%%%%%%%%%%%%%%%%%%%%%%%%%%%%%%%

The selected $e'p'\gamma$ event sample was divided into 166 four-dimensional kinematical bins, with 5 bins in the $Q^2$-$x_B$ space, 4 in $-t$, and 10 in $\phi$, and according to the sign of the target polarization with respect to the beam direction. %The choice of the binning was optimized according to the available statistics of the data, in order to achieve as much an even population of events per bin as possible. %Bins with less than 5 events per beam-target polarization state were discarded resulting in 165 retained bins. 
Asymmetries were then reconstructed for each bin according to
\begin{equation}\label{eq_asym}
A_{UL} = \frac{1}{D_f} \frac{ (N_{+} - N_{-} )}{(N_{+} P_{t}^{-}+ N_{-}P_{t}^{+} )},
\end{equation}
where $N_{+(-)}$ are the number of counts, normalized by the accumulated charge measured by the Faraday cup for each target-polarization sign and $P_{t}^{+(-)}$ are the values of the positive (negative) target polarizations. 

For each bin and for each target polarization sign, the counts $N$ were corrected by subtracting from the $e'p'\gamma$ yield the contamination from $e'p'\pi^0$ events in which one of the two $\pi^0$-decay photons had escaped detection. The contamination %($N_{\pi^01\gamma}$) 
was computed as the product of the yield for measured $e'p'\pi^0$ events times the ratio of the acceptances, obtained via Monte-Carlo simulations, for $e'p'\pi^0$ events applying, respectively, the $e'p'\gamma$ selection cuts and the cuts needed to select the $e'p'\pi^0$ final state.
The average effect of the background subtraction was $\sim 11\%$ of the asymmetry at $90^{\circ}$, and typically smaller than the statistical uncertainties. 

The target polarizations $P_t^{\pm}$ ($P_t^+\simeq 80\%$, $P_t^-\simeq 74\%$) were computed by extracting the product of beam and target polarizations ($P_{b}P_{t}$) measuring the well-known elastic-scattering asymmetry \cite{donnelly} and using the beam polarization value ($P_{b}\simeq 84\%$) that had been measured during dedicated M{\o}ller runs throughout the experiment. 

The asymmetry, which in Eq. \ref{eq_asym} is defined with respect to the beam direction to which the target polarization was aligned, was corrected to be redefined with respect to the virtual photon, thus becoming consistent with the convention adopted in most theoretical calculations. 
On average this correction modifies the asymmetry by 4\% relative to its value at $90^{\circ}$, which is always much smaller than the statistical uncertainties. The same holds for its associated systematic uncertainty. Bin-centering corrections, which had minimal impact, were also applied. 

The main source of systematic uncertainties is the sensitivity of the results to the  exclusivity cuts. Other sources of systematic uncertainties are the dilution factor, the beam and target polarizations, and the $\pi^{0}$ contamination. These effects were estimated on a bin-by-bin basis, recomputing the asymmetry varying within reasonable limits each factor of uncertainty. The individual systematic uncertainties were then added in quadrature, and their average, relative to the average value of the asymmetry at $90^{\circ}$, is $\sim 15\%$. For 97\% of the data points the total systematic uncertainty was found to be smaller than the statistical uncertainty. 

\begin{figure} 
\includegraphics[width = 8.6 cm]{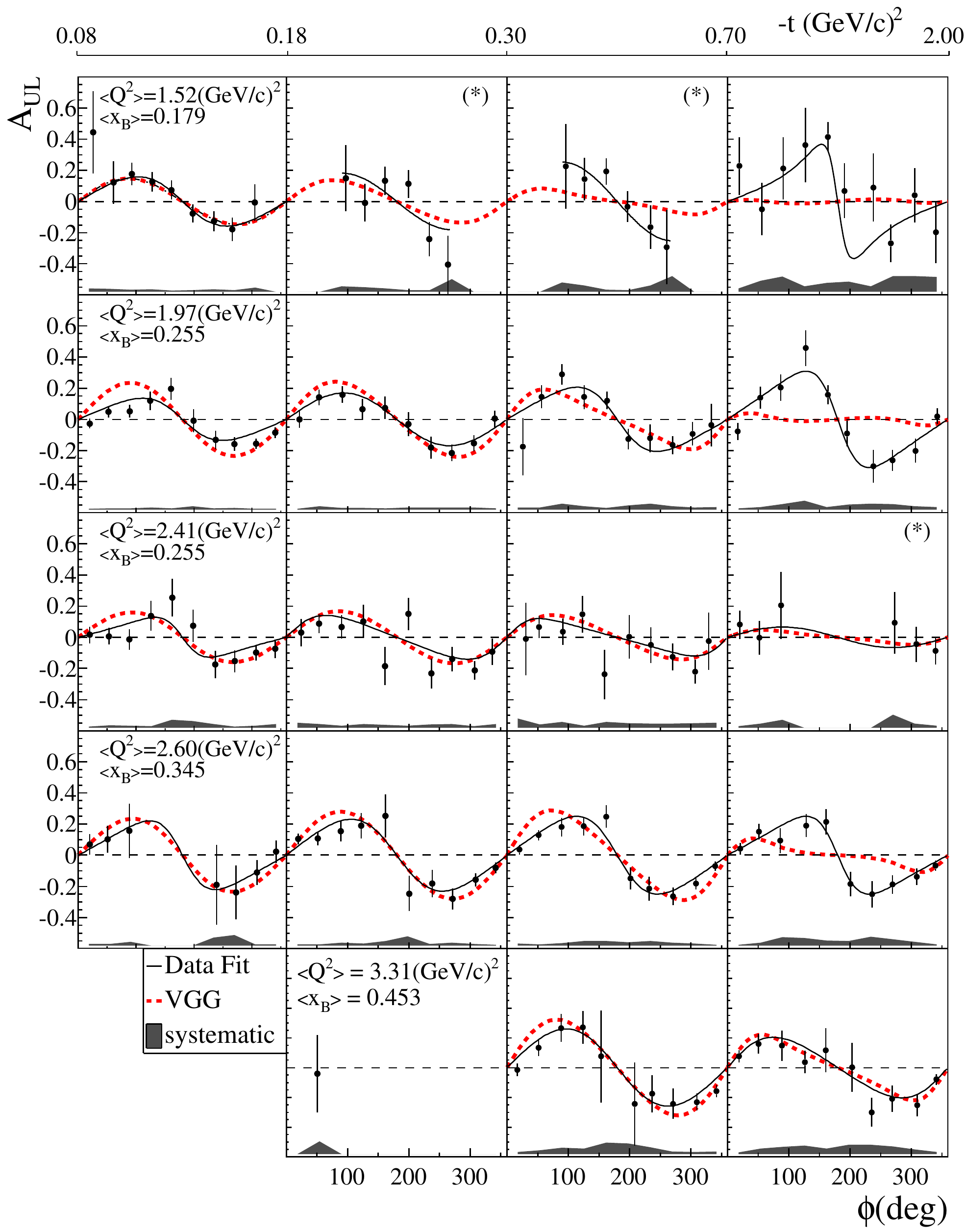}
\vspace{-0.8cm}
\caption{\label{fig3} (Color online) Target-spin asymmetry ($A_{UL}$) for DVCS/BH events plotted as a function of $\phi$ for each three-dimensional bin in $Q^2$-$x_B$ (rows) and $-t$ (columns - the bin limits are shown on the top axis). The shaded bands are the systematic uncertainties. The thin black line is the fit to $A_{UL}$ with the function $\frac{\alpha \sin\phi}{1+\beta\cos\phi}$ (for all bins but those marked with $(*)$, which were fitted with $\alpha\sin\phi$ due to the limited $\phi$ coverage). The dashed/red lines are the predictions of the VGG model \cite{vgg}.}
\vspace{-0.4cm}
\end{figure}

The resulting target-spin asymmetries, covering the kinematic ranges $1<Q^2<5.2$ (GeV/c)$^2$, $0.12<x_B<0.6$, $0.08 <-t<2$ (GeV/c)$^2$, are shown as a function of $\phi$ in Fig.~\ref{fig3}. 
The asymmetries exhibit a clear $\sin\phi$-type modulation, which is expected at leading twist for the interference of DVCS and BH. The average amplitude is $\sim 0.2$. The variable for which the biggest variations in shape and amplitude are observed is $-t$.

\begin{figure} 
\includegraphics[width = 8.6 cm]{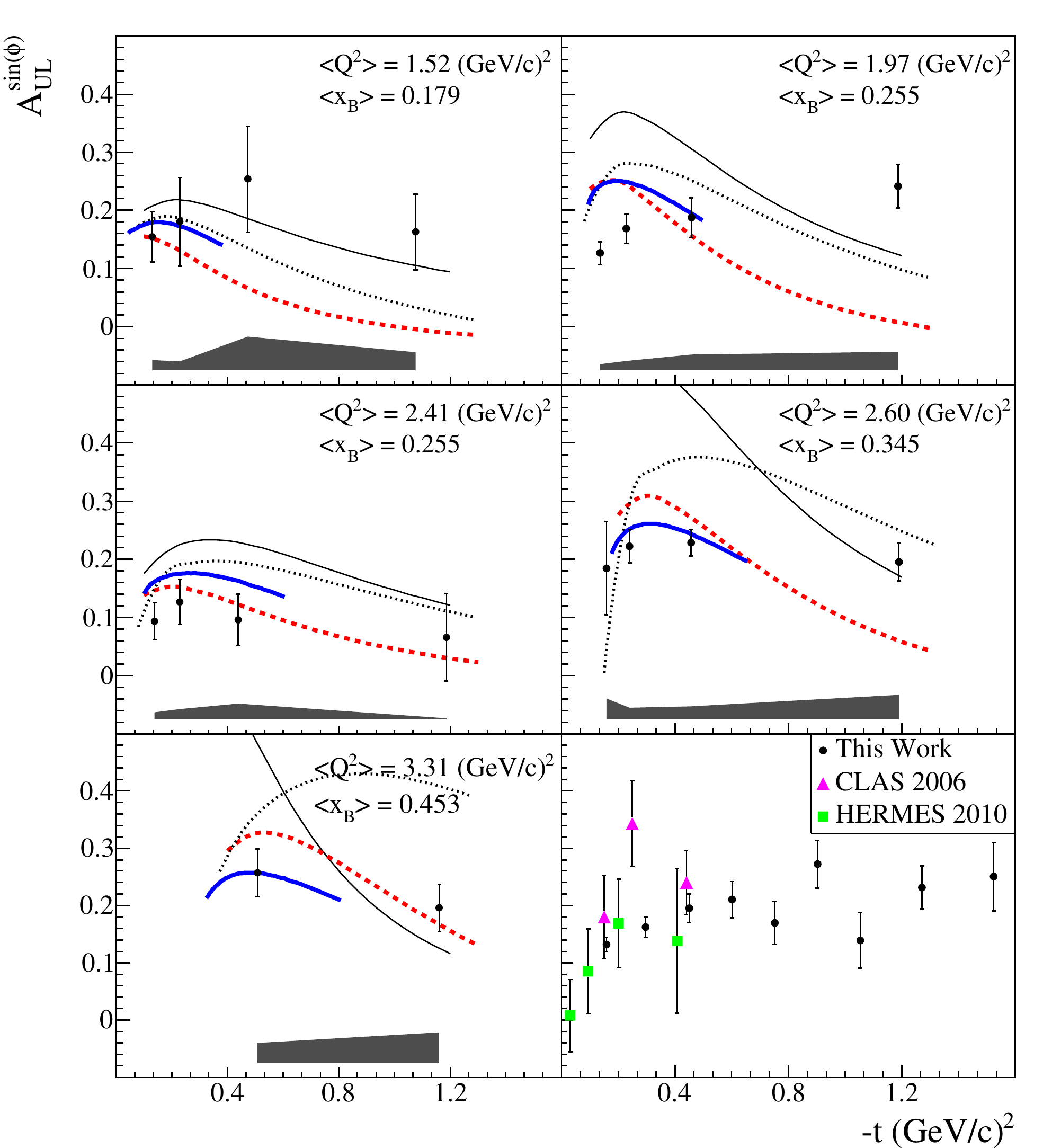}
\vspace{-0.7cm}
\caption{\label{fig4} (Color online) First five plots: $-t$ dependence of the $\sin\phi$ amplitude of $A_{UL}$ for each $Q^2$-$x_B$ bin. The shaded bands represent the systematic uncertainties. The curves show the predictions of four GPD models: i) VGG \cite{vgg} (red-dashed), ii) GK \cite{kroll} (black-dotted), KMM12 \cite{kreso} (blue-thick solid), GGL \cite{liuti} (black-solid). Bottom right plot: comparison of the $\sin\phi$ amplitude of $A_{UL}$ as a function of $-t$ for the results of this work (black dots) integrated over all $Q^2$ and $x_B$ values ($\langle Q^2 \rangle=2.4$ (GeV/c)$^2$, $\langle x_B\rangle=0.31$), the HERMES results \cite{hermes} (green squares) at $\langle Q^2 \rangle=2.459$ (GeV/c)$^2$, $\langle x_B\rangle=0.096$, and the previously published CLAS results \cite{shifeng} (pink triangles), at $\langle Q^2\rangle=1.82$ (GeV/c)$^2$, $\langle x_B \rangle=0.28$. }
\vspace{-0.4cm}
\end{figure}

The measured $\phi$ distributions of $A_{UL}$ were fit, where possible, with the function of Eq.~\ref{eq_aUL}. Fits were also done, where the statistics allowed, adding an extra $\sin 2\phi$ term to the numerator. This term turned out to be negligible compared to the $\sin\phi$ term, gaining strength in the low-$Q^2$ kinematics as $-t$ increased. Interpreted in the GPD framework, this result points towards the dominance of the leading-twist handbag process of Fig.~\ref{fig1} over higher-twist diagrams. The $-t$ dependence of the $A_{UL}$ fit parameter $\alpha$ for each bin in $Q^2$-$x_B$ is shown in Fig.~\ref{fig4}, panels 1-5. The systematic uncertainties on $\alpha$ are represented by the dark shaded bands. The trend of the target-spin asymmetry as a function of $-t$ is quite different from what was observed for the beam-spin asymmetry \cite{fx}, which displayed a much stronger drop, by about a factor of 3 on average, for all $Q^2$-$x_B$ kinematics but more markedly at low $x_B$, as $-t$ is increased over the same range as this measurement. It must be recalled that the DVCS/BH beam-spin and target-spin asymmetries are mostly sensitive to the GPDs $H$ and to a combination of $\tilde{H}$ and $H$, respectively. Therefore, considering that the $t$-slope of the GPDs is linked via a Fourier-like transform to the transverse position of the struck parton, this result suggests that the axial charge (linked to $\Im{\rm m}{\tilde{\cal H}}$) is more concentrated in the center of the nucleon than the electric charge (linked to $\Im{\rm m}{\cal H}$), confirming what was first observed in \cite{fitmick}. This is in agreement with the behavior as a function of $Q^2$ of the axial form factor, which is the first moment in $x$ of $\widetilde{H}$, and which was measured in $\pi^+$ electroproduction experiments on the proton as well as in neutrino-nucleon scattering \cite{bodek}. Our result adds to this the extra information on the longitudinal momentum of the partons.

The sixth panel of Fig.~\ref{fig4} shows our comparison of $A_{UL}$ with the previous world data from HERMES \cite{hermes} and CLAS \cite{shifeng}: here our data were integrated over $Q^2$-$x_B$, as there is no overlap between our 5 bin centers and the central kinematics of the other datasets, and were fitted for 9 intervals in $-t$ with the function $\alpha\sin\phi + \beta\sin 2\phi$ to be consistent with the fits employed for the other data. Our results, in agreement with the previous ones, improve the existing statistics by more than a factor of 5 in the $-t$ region up to $\sim 0.4$ (GeV/c)$^2$, and extend the $-t$ range up to 1.6 (GeV/c)$^2$. 

In panels 1-5 of Fig.~\ref{fig4} predictions from four GPD-based models, listed in the caption, are included. Both the VGG and GK models are based on double distributions \cite{radyuskin2,muller} to parametrize the $(x,\xi)$ dependence of the GPDs, and on Regge phenomenology for their $t$ dependence. The main differences between these two models are in the parametrization of the high-$t$ part of the electromagnetic form factors and in the fact that the parameters of the GK model are tuned using low-$x_B$ deeply-virtual meson production data from HERA. KMM12 is a hybrid model designed for global fitting, in which sea-quark GPDs are represented as infinite sums of $t$-channel exchanges; valence quarks are modeled in terms of these GPDs on the line $\xi = x$. The parameters of KMM12 were fixed using polarized- and unpolarized-proton DVCS data from HERMES \cite{hermes_bsa2,hermes}. The kinematic range of applicability of this model is defined by the relation $-t<\frac{Q^2}{4}$. The GGL model provides a diquark-model inspired parametrization of the GPDs that incorporates Regge behavior for the $t$ dependence. The GGL model parameters were obtained by fitting both DIS structure functions and the recent flavor-separated nucleon form factor data \cite{cates}. 

While the VGG and GK models are in fair agreement with the data at low $-t$, especially for the lowest $Q^2$-$x_B$ bin, the quark-diquark-based model tends to diverge away from the measured $A_{UL}$ values going toward higher $x_B$ for all $-t$. The data do not exhibit as strong a drop at high $-t$ as the four models predict. In the low-$Q^2$ and high-$t$ region, where we also observe a change of shape in the $\phi$ distribution compared to the model predictions (see Fig.~\ref{fig3}, last columns of the first two rows), the leading-twist approximation, which is at the core of all these GPD models, could be one of the causes of the discrepancies. 
The predictions of the VGG and GK models are, as expected, quite similar, as they share common concepts, but start to differ as $x_B$ increases: this is to be expected because the GK model contains parameters that were tuned using low-$x_B$ HERA data on meson production, and therefore it is not optimized for the valence region ($x_B \gtrapprox  0.3$). Moreover, the parametrization of the $-t$ dependence, although Regge-inspired in both cases, is handled differently in the two models.
The KMM12 model gives the best fit to the data, especially at the highest $x_B$, but due to its $-t<\frac{Q^2}{4}$ prescription it cannot be applied to all the available kinematic bins. 

In summary, for the first time four-dimensional target-spin asymmetries with longitudinally polarized protons arising from the interference of deeply virtual Compton scattering and Bethe-Heitler were extracted over a large phase space. $A_{UL}$ was measured for 166 bins in $Q^2$, $x_B$, $-t$ and $\phi$, with an average statistical precision of $\sim25\%$, which largely dominates the systematic uncertainties. The $\phi$ dependence of the obtained asymmetries was studied. Interpreting this result in the GPD framework, the dominance of the leading-twist handbag mechanism can be observed via the prevalence of the $\sin\phi$ term, especially at low $t$ and high $Q^2$. The $t$ slope of the asymmetry, shallower with respect to that of the beam-spin asymmetry in the same kinematic range, suggests, within the leading-twist approximation, that the axial charge is more focused in the center of the proton than the electric charge. Predictions of four GPD-based models are in qualitative agreement at low $Q^2$-$x_B$ and $-t$ with the data, but fail to predict the correct $t$ dependence of the data in the other kinematics, proving the importance of our results to improve the parametrizations of the GPD $\widetilde{H}$. Thanks to their vast $t$ coverage, our results can also provide a starting point to understand higher-twist effects. These data, combined with the beam-spin asymmetry results from CLAS \cite{fx} and with the DVCS/BH double-spin asymmetry obtained using this same data set \cite{silviap}, will bring strong constraints for model-independent extractions of Generalized Parton Distributions \cite{fitmick2,fitherve,mick_herve,fitkreso}.

We thank the staff of the Accelerator and Physics Divisions and of the Target Group at Jefferson Lab for making the experiment possible. Special thanks to M. Guidal, F. Sabati\'e, S. Liuti, D. M\"uller and K. Kumericki for the theoretical support. This work was supported in part by the U.S. Department of Energy (No. DE-FG02-96ER40950) and National Science Foundation, the French Centre National de la Recherche Scientifique and Commissariat  \`a l'Energie Atomique, the French-American Cultural Exchange (FACE), the Italian Istituto Nazionale di Fisica Nucleare, the Chilean Comisi\'on Nacional de Investigaci\'on Cient\'ifica y Tecnol\'ogica (CONICYT), the National Research Foundation of Korea, and the UK Science and Technology Facilities Council (STFC). The Jefferson Science Associates (JSA) operates the Thomas Jefferson National Accelerator Facility for the United States Department of Energy under contract DE-AC05-06OR23177.

\end{document}